\documentclass[a4paper,twoside]{article}
\usepackage{amsmath,amssymb,amsfonts}
\usepackage{graphics}                 
\def\R{{\mathbb R}}

\date{\small\it Draft \today}
\title{Quantifying galactic clustering and departures from randomness of the inter-galactic
void probability function using information geometry}

\author{C.T.J. Dodson, \\ School of Mathematics,
\\ University of Manchester, Manchester M60 1QD, UK \\ ctdodson@manchester.ac.uk}

\begin{document}
\maketitle
\begin{abstract}
A number of recent studies have estimated the inter-galactic void
probability function and investigated its departure from various
random models. We study a family of parametric statistical models
based on gamma distributions, which do give realistic descriptions
for other stochastic porous media. Gamma distributions contain as
a special case the exponential distributions, which correspond to
the `random' void size probability arising from Poisson processes.
 The space of
parameters is a surface with a natural Riemannian metric
structure. This surface contains the Poisson processes as an
isometric embedding and a recent theorem~\cite{gamran} shows that
it contains neighbourhoods of all departures from randomness. The
method provides thereby a geometric setting for quantifying
departures from randomness and on which may be formulated
cosmological evolutionary dynamics for galactic clustering and for
the concomitant development of the void size distribution.

The 2dFGRS data~\cite{croton04a,croton04b} offer the possibility of more detailed
investigation of this approach than was possible when it was
originally suggested~\cite{vpf,ijtp,gsis} and some parameter estimations are given.

\end{abstract}

\section{Introduction}
Several years ago, the author presented an information
geometric approach to modelling a space of perturbations of the
random state for galactic clustering and cosmological void
statistics~\cite{vpf}.  The present note is intended to update somewhat and draw attention
to this approach as a possible contribution to the interpretation of the data from
the 2-degree field Galaxy Redshift Survey (2dFGRS),
cf Croton et al.~\cite{croton04a,croton04b}.

The classical random model is that arising
from a Poisson process of mean density $\overline{N}$ galaxies per unit
volume in a large box. Then, in a region of volume
$V,$ the probability of finding exactly $m$ galaxies is
\begin{equation}
P_m =\frac{(\overline{N}V)^m}{m!} e^{-\overline{N}V} \label{poisson}
\end{equation}
So the probability that the given region is devoid of galaxies is
$P_0=e^{-\overline{N}V}.$ It follows that the probability density function
for the continuous random variable $V$ in the Poisson case is
\begin{equation}
p_{random}(V) = \overline{N} \, e^{-\overline{N}V} \label{negexp}
\end{equation}
In practice of course, measurements will depend on algorithms that specify
threshold values for density ranges of galaxies in cells and the lowest
range will represent the `underdense' regions which include the voids;
Benson et al.~\cite{benson} discuss this.

A hierachy of $N$-point correlation functions needed to represent
clustering of galaxies in a complete sense was devised by
White~\cite{white} and he provided explicit formulae, including
their continuous limit. In particular, he made a detailed study of
the probability that a sphere of radius $R$ is empty and showed
that formally it is symmetrically dependent on the whole hierarchy
of correlation functions. However, White concentrated his
applications on the case when the underlying galaxy distribution
was a Poisson process, the starting point for the present approach
which is concerned with geometrizing the parameter space of
departures from a Poisson process. Croton et al.~\cite{croton04b} found that
the negative binomial model for galaxy clustering gave a very good
approximation to the 2dFGRS, pointing out that this model is a discrete
version of the gamma distribution.

\section{Modelling statistics of galaxy void sizes}
For a general account of large-scale structures in the universe,
see Fairall~\cite{fairall}.
Kauffmann and Fairall~\cite{kauffmannfairall} developed a catalogue
search algorithm for larger nearly spherical regions devoid of bright galaxies
and obtained a spectrum for radii of significant voids. This indicated
a peak radius near 4 $h^{-1}Mpc,$ a long tail stretching at least to
32 $h^{-1}Mpc,$ and is compatible with the recent extrapolation models
of Baccigalupi et al~\cite{bacc} which yield an upper bound on void radii
of about 50 $h^{-1}Mpc.$ This data has of course omitted the expected very
large numerical contribution of smaller voids. More recent work, notably of
Croton et al.~\cite{croton04b} provide much larger samples with improved
estimates of void size statistics and Benson et al.~\cite{benson} gave a
theoretical analysis in anticipation of the 2dFGRS survey data, including
the evaluation of the void and underdense probability functions. Hoyle and Vogeley~\cite{hoyle}
provided detailed results for the statistics of voids larger than 10 $h^{-1}Mpc$
in the 2dFGRS survey data; they concluded that
such voids constitute some $40\%$ of the universe and
have a mean radius of about 15 $h^{-1}Mpc.$

The count density $N(V)$ of galaxies observed in zones using a range of sampling schemes
each with a fixed zone volume $V$ results in a decreasing variance $Var(N(V))$
of count density with increasing zone size, roughly of the form
\begin{eqnarray}
  Var(N(V)) &\approx& V(0)\, e^{-V/V_k} \ \ {\rm as} \ V\rightarrow 0
\end{eqnarray}
where $V_k$ is some characteristic scaling parameter. This monotonic decay of variance
with zone size is a natural consequence of the monotonic decay of the covariance function,
roughly isotropically and of the form
\begin{eqnarray}
  Cov(r) &\approx&  e^{-r/r_k} \ \ {\rm as} \ r\rightarrow 0
\end{eqnarray}
where $r_k$ is some characteristic scaling parameter of the order of magnitude
of the diameter of filament structures; this was discussed in~\cite{ijtp}. Then
\begin{eqnarray}
  Var(N(V)) &\approx& \int_0^\infty Cov(r) \, b(r) \, dr
\end{eqnarray}
where $b(r)$ is the probability density of finding two points separated
by distance $r$ independently and at random in a zone of volume $V.$ The power spectrum
using, say, cubical cells of side lengths $R$ is given by the family of integrals
\begin{eqnarray}
  Pow(N(R)) &\approx& \int_R^\infty Cov(r) \, b(r) \, dr.
\end{eqnarray}

Fairall~\cite{fairall} (page 124) reported a value $\sigma^2=0.25$
for the ratio of variance $Var(N(1))$ to mean squared $\overline{N}^2$
for counts of galaxies in cubical cells of unit side length. In other words, the coefficient
of variation for sampling with cells of unit volume is
\begin{equation}
cv(N(1))=\frac{\sqrt{Var(N(1))}}{\overline{N}}=0.5
\end{equation}
and this is dimensionless.

We choose a family of parametric statistical models for void volumes that
includes the random model~(\ref{negexp}) as a special case. There are of course
many such families, but we take one that a recent
theorem~\cite{gamran} has shown contains neighbourhoods of all
departures from randomness and it has been successful in modelling
void size distributions in terrestrial stochastic porous
media with similar departures from randomness~\cite{dodsonsampson}.
Also, the complementary logarithmic version has been
used in the representation of clustering of
galaxies~\cite{ijtp,gsis}. The family of gamma distributions has
event space $\Omega=\R^+,$ parameters $\mu,\beta\in \R^+$ and
probability density functions given by
\begin{equation}
f(V;\mu,\beta)=\left(\frac{\beta}{\mu}\right)^\beta \,
\frac{V^{\beta-1}}{\Gamma(\beta)}\, e^{-V\beta/\mu}
\label{feq}
\end{equation}
Then $\overline{V}=\mu$ and $Var(V)=\mu^2/\beta$ and we see that $\mu$ controls
the mean of the distribution while the spread and shape is controlled
by, $1/\beta,$ the square of the coefficient
of variation.

The special case $\beta=1$ corresponds to the situation when $V$
represents the random or Poisson process in~(\ref{negexp}) with
$\mu=1/n.$ Thus, the family of gamma distributions can model a
range of stochastic processes corresponding to non-independent
`clumped' events, for $\beta<1,$ and dispersed events, for
$\beta>1,$ as well as the random case
(cf.~\cite{gamran,InfoGeom,dodsonsampson}). Thus, if we think of this range
of processes as corresponding to the possible distributions of
centroids of extended objects such as galaxies that are initially
distributed according to a Poisson process with $\beta=1,$ then
the three possibilities are:
\begin{description}
\item[Chaotic or random structure] with no interactions among
constituents, $\beta=1;$
\item[Clustered structure] arising from mutually attractive type
interactions, $\beta<1;$
\item[Dispersed structure] arising from mutually repulsive type
interactions, $\beta>1.$
\end{description}

For our gamma-based void model we consider the radius $R$ of a spherical void with
volume $V=\frac{4}{3} \pi R^3$ having distribution~(\ref{feq}).
Then the probability density function for $R$ is given by
\begin{equation}
p(R;\mu,\beta)= \frac{4\pi R^2}{\Gamma(\beta)}\, \left(\frac{  \beta }{
\mu }\right)^\beta \, \left(\frac{4 \pi R^3}{3}\right)^{\beta-1}
\, e^\frac{ -4\pi R^3\beta}{3\mu} \label{Rpdf}
\end{equation}
The mean $\overline{R},$ variance $Var(R)$ and coefficient of
variation $cv(R)$ of $R$ are given, respectively, by
\begin{eqnarray}
\overline{R}&=& \left(\frac{3\mu}{4\pi\beta}\right)^\frac{1}{3}\,
\frac{\Gamma(\beta+\frac{1}{3})}{\Gamma(\beta)}
\label{meanR} \\
Var(R)&=& \left(\frac{3\mu}{4\pi\beta}\right)^\frac{2}{3}
\frac{
    \Gamma(\beta) \ \Gamma(\beta+\frac{2}{3})-\Gamma(\beta+\frac{1}{3})^2 }{\Gamma(\beta)^2}
\label{varR} \\
cv(R)&=&\frac{\sqrt{Var(R)}}{\overline{R}}  =
\sqrt{\frac{\Gamma (\beta )\  \Gamma \left(\beta+\frac{2}{3}\right)}
{\Gamma \left(\beta+\frac{1}{3}\right)}-1}      \label{cvR}
\end{eqnarray}

The fact that the coefficient of variation~(\ref{cvR}) depends
only on $\beta$ gives a rapid parameter fitting of data to the probability
density function for void radii~(\ref{Rpdf}).
Numerical fitting to~(\ref{cvR}) gives $\beta;$ this substituted
in~(\ref{meanR}) yields an estimate of $\mu$ to fit a given
observational mean.

However, there is a complication: necessarily in order to have a physically
meaningful definition for voids, observational measurements introduce
a minimum threshold size for voids.
For example, Hoyle and Vogeley~\cite{hoyle} used algorithms to obtain statistics on 2dFGRS
voids with radius $R>R_{min} = 10 \ h^{-1}Mpc;$ for voids above this threshold
they found their mean size is about $15 \ h^{-1}Mpc$ with a variance of about $8.1.$ This of
course is not directly comparable with the above distribution for $R$ in equation (\ref{Rpdf})
since the latter has domain $R>0.$ Now, from (\ref{Rpdf}), the probability that a void has
radius $R>A$ is
\begin{equation}
P_{A}=\frac{ \Gamma (\beta ,\frac{4 A^3 \pi
   \beta }{3 \mu }) } {\Gamma (\beta )} \label{PgeqA}
   \end{equation}
and hence the mean, variance and coefficient of
variation for the void distribution with $R>A$ become:
\begin{eqnarray}
  \overline{R_{>A}} &=& \left(\frac{3\mu}{4\pi\beta }\right)^{\frac{1}{3}}\
   \frac{
  \Gamma(\beta +\frac{1}{3},\frac{4 A^3 \pi  \beta }{3 \mu })}
  {\Gamma (\beta ,\frac{4 A^3 \pi
   \beta }{3 \mu })}    \label{RgAbar}\\
  Var(R_{>A}) &=& \left(\frac{3\mu}{4\pi\beta }\right)^{\frac{2}{3}} \ \frac{
  \Gamma(\beta )\ \Gamma (\beta +\frac{2}{3},\frac{4 A^3 \pi  \beta
   }{3 \mu })-\Gamma (\beta +\frac{1}{3},\frac{4 A^3 \pi
    \beta }{3 \mu })^2}{
   \Gamma (\beta ,\frac{4 A^3 \pi
   \beta }{3 \mu }) \ \Gamma (\beta )}  \label{RgAVar}\\
  cv(R_{>A}) &=& \frac{\sqrt{Var(R_{>A})}}{\overline{R_{>A}}}  =
  \sqrt{\frac{\Gamma (\beta )\ \Gamma (\beta +\frac{2}{3},\frac{4 A^3
   \pi  \beta }{3 \mu })}{\Gamma (\beta
   +\frac{1}{3},\frac{4 A^3 \pi  \beta }{3 \mu })^2}-1} \label{RgAcv}
\end{eqnarray}
where
$$\Gamma(\beta,A)= \int_A^\infty t^{\beta-1} \, e^{-t} \, dt \ \
 {\rm is \ the \ incomplete \ gamma \ function \ with} \ \Gamma(\beta)=\Gamma(\beta,0).$$

Summarizing from Hoyle and Vogeley~\cite{hoyle}:\\
$A = 10 \  h^{-1}Mpc,$ $P_{A}\approx 0.4,$
$\overline{R_{>A}}\approx 15 \ h^{-1}Mpc,$ $Var(R_{>A})\approx 8.1 \ (h^{-1}Mpc)^2$ so
$cv(R_{>A})\approx 0.19.$

\begin{figure}
\begin{center}
\begin{picture}(300,260)(0,0)
\put(-50,-10){\resizebox{10cm}{!}{\includegraphics{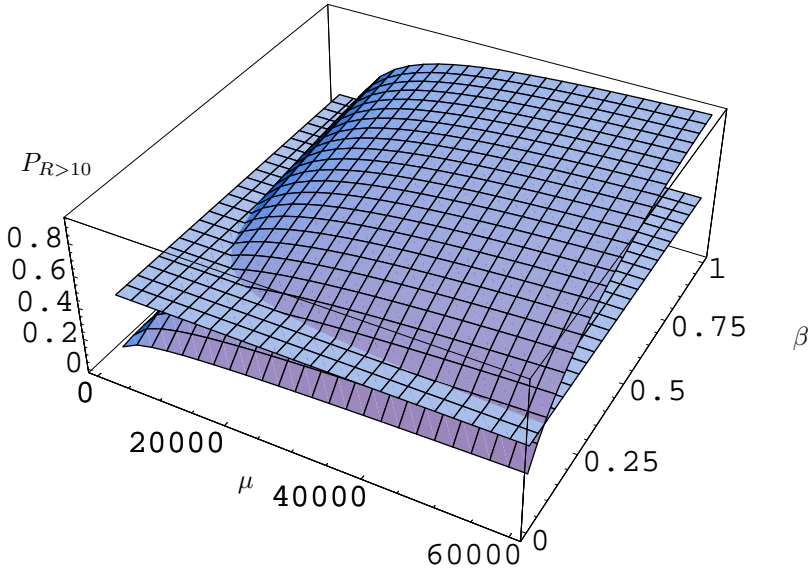}}}
\put(40,30){$\mu$}
\put(-42,150){$P_{R>10}$}
\put(250,84){$\beta$}
\end{picture}
\end{center}
\caption{{\em Probability that a void will have radius $R>10 \  h^{-1}Mpc$ as a function
of parameters $\mu, \, \beta$ from equation (\ref{PgeqA}).
The range $\beta<1$ corresponds to clustering regimes.
The plane at level $P_{R>10}=0.4$ corresponds to the fraction $40\%$ of the universe
filled by voids, as reported by Hoyle and Vogeley~\cite{hoyle}.}}
\label{PRgeqA}
\end{figure}

\section{Model coupling of clustering densities and voids}
Next we follow the methodology introduced in~\cite{ijtp,gsis} to provide a model that
links the number counts in cells and the void probability function and which contains
perturbations of the random case. This exploits the central role of the gamma distribution
in providing neighbourhoods of randomness~\cite{gamran} that contain all maximum
likelihood nearby
perturbations of the random case and it allows a direct use of the linked information
geometries for the coupled stochastic processes of voids and galaxies.
Clearly, in regions where the local void volume $V$ tends to be small the local matter
density $N$ will tend to be large. Since the matter density must be bounded, then a simple
phenomenological model that couples the two random variables in the stochastic process
is $N(V)=e^{-V},$ where the upper bound on $N$ has been set to unity. This model was
explored in~\cite{ijtp,gsis} and it is easy to show that the probability density function
for $N$ is given by the log gamma distribution
\begin{equation}
g(N;\mu,\beta)=\left(\frac{\beta}{\mu}\right)^\beta \,
\frac{N^{\beta/\mu -1}}{\Gamma(\beta)}\, |\log{N}|^{1-\beta}
\label{geq}
\end{equation}
This distribution for local galactic number density has mean $\overline{N},$
 variance $Var(N)$ and coefficient of variation $cv(N)=\sqrt{Var(N)}/\overline{N}$ given by
\begin{eqnarray}
\overline{N}&=& \left(\frac{\beta}{\beta+\mu}\right)^\beta \\ Var(N)&=&
\left(\frac{\beta}{\beta+2\mu}\right)^\beta -
\left(\frac{\beta}{\beta+\mu}\right)^{2\beta}\\
cv(N)=\frac{\sqrt{Var(N)}}{\overline{N}} &=&
\sqrt{\left(\frac{\beta }{\beta +\mu }\right)^{-2 \beta } \left(\frac{\beta }{\beta +2 \mu
   }\right)^{\beta }-1}.\label{cvN}
\end{eqnarray}
Using the reported value $cv(N(1))=0.5$ from Fairall~\cite{fairall} (page 124)
for cubical volumes with side length $R=1 \ h^{-1}Mpc,$ the curve
so defined in the parameter space for the log gamma distributions (\ref{geq}),
has maximum clustering for $\beta\approx 0.6, \mu\approx 0.72.$

From the 2-degree field Galaxy Redshift Survey (2dFGRS),
Croton et al.~\cite{croton04a} in Figure 2 reported the decay of normalised variance,
$\overline{\xi_2}=cv(N(R))^2$ with scale radius $R$ and the associated departure
from randomness in the form $\chi=-\frac{\log_{10}P_0(R)}{\overline{N}},$ where
$P_0(R)$ is the probability of finding zero galaxies in a spherical region of radius $R$
when the mean number is $\overline{N}.$
From that Figure 2 we see that, for the data of the Volume Limited Catalogue with magnitude
range $-20$ to $-21$ and $\overline{N}(1)=1.46:$
$cv(N(1))^2\approx 6$ and $\chi\approx 0.9$ at $R\approx 1$ also
$cv(N(7))^2\approx 1$ and $\chi\approx 0.4$ at $R\approx 7.$

Croton et al.~\cite{croton04b} in Table 1 reported
$\overline{N}$ values for cubical volumes with side length $R=1 \ h^{-1}Mpc$ in the
range $0.11\leq \overline{N} \leq 11.$ From Figure 3 in that paper we see that,
at the scale $R=1 \ h^{-1}Mpc,$ $\log_{10}\overline{\xi}_2\approx 1$ which gives a
coefficient of variation  $cv(N(1)) \approx \sqrt{10}.$

\begin{figure}
\begin{center}
\begin{picture}(300,260)(0,0)
\put(-50,-10){\resizebox{10cm}{!}{\includegraphics{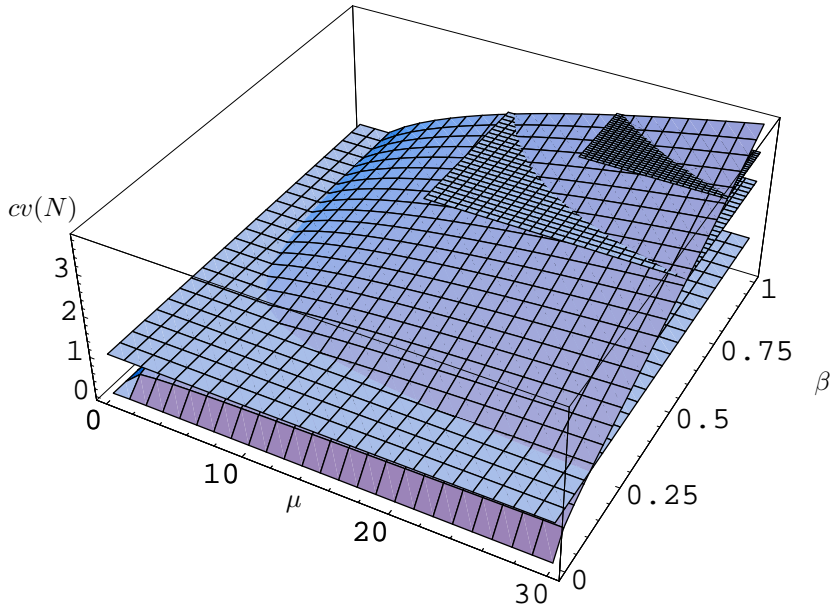}}}
\put(40,30){$\mu$}
\put(-65,140){$cv(N)$}
\put(240,74){$\beta$}
\end{picture}
\end{center}
\caption{{\em Coefficient of variation of counts in cells $N$ for log gamma distribution
equation~\protect(\ref{geq}). The range $\beta<1$ corresponds to clustering regimes.
The three planes show the levels $cv(N)=1, \, \sqrt{6}, \, \sqrt{10}$
as reported by Fairall~\cite{fairall} and Croton et al.~\cite{croton04a,croton04b}.}}
\label{cvN1610}
\end{figure}

The above-mentioned observations $cv(N)=1, \, \sqrt{6}, \, \sqrt{10}$
are shown in Figure~\ref{cvN1610} on a plot of the coefficient of variation for
the number counts in cells from the log gamma family of distributions
equation~\protect(\ref{geq}). The range $\beta<1$ corresponds to clustering regimes.

Theoretical models for the evolution of galactic clustering through an
evolving stochastic process subordinate to the log gamma distribution (\ref{geq})
of densities could be
represented as curves on the space of parameters with the metric (\ref{gammametric})
 interpreting the parameter changes with time in the appropriate way. The coupling
 with the void probability function controlled by the gamma distribution (\ref{Rpdf})
allows the corresponding void evolution to be represented. It is of course very
unlikely that this simple model is suitable in all respects but
given a different family of distributions the necessary information geometry
can be computed for the representation of evolutionary processes.

\section{Information geometry of gamma models for void volume statistics}
For stochastic processes
subordinate to a given family of parametric statistical models,
Shannon's information theoretic `entropy' or `uncertainty'
(cf. eg. Jaynes~\cite{jaynes}) is given, up to a factor, by the
negative of the expectation of the logarithm of the probability density function.
For the family of models we propose for
void volumes~(\ref{feq}) this entropy is,
\begin{eqnarray}
S_f(\mu,\beta) &=& -\int_0^\infty \log(f(V;\mu,\beta)\, f(V;\mu,\beta) \, dV \\
               &=& \beta+(1-\beta)\frac{\Gamma'(\beta)}{\Gamma(\beta)} +
\log\frac{\mu \, \Gamma(\beta)}{\beta} \label{fentropy}
\end{eqnarray}
In particular, at unit mean, the maximum entropy (or maximum uncertainty)
occurs at $\beta=1,$ which is the random case, and then $S_f(\mu,1)=1+\log\mu.$

The `maximum likelihood' estimates $\hat{\mu}, \hat{\beta}$ of $\mu,\beta$
can be expressed in terms of the mean and mean logarithm of a
set of independent observations $X=\{X_1,X_2,\ldots,X_n\}.$ These estimates are obtained
in terms of the properties of $X$ by maximizing the `log-likelihood' function
$$l_X(\mu,\beta)=\log lik_X(\mu,\beta)=\log\left(\prod_{i=1}^np(X_i;\mu,\beta)\right)$$
with the following result
\begin{eqnarray}
\hat{\mu}&=&\bar{X}=\frac{1}{n}\sum^n_{i=1}X_i\\ \log\hat{\beta}
-\psi(\hat{\beta})
   & = & \overline{\log X} - \log\bar{X}
\end{eqnarray}
where $\overline{\log X}=\frac{1}{n}\sum^n_{i=1}\log X_i$ and
$\psi(\beta)=\frac{\Gamma'(\beta)}{\Gamma(\beta)}$ is the digamma
function, the logarithmic derivative of the gamma function

The usual Riemannian information metric on the 2-dimensional parameter
space $\cal{S}=\{(\mu,\beta)\in\R^+\times\R^+\}$
is given by
\begin{equation}
ds_{\cal{S}}^2=\frac{\beta}{\mu^2} \, d\mu^2 +
        \left(\psi'(\beta) -\frac{1}{\beta}\right)\, d\beta^2 \ \
        \ {\rm for} \ \mu,\beta\in\R^+ . \label{gammametric}
\end{equation}
The important point about this non-Euclidean metric on the space of parameters
is that it derives from log-likelihood properties and so it is the `correct'
one for this family of distributions. Given a different family then the
information geometry can be computed for that; the geometries are known
also for bivariate gamma distributions, Gaussian and multivariate Gaussian
distributions among others. For more details about the geometry
see~\cite{InfoGeom}.

The 1-dimensional subspace
parametrized by $\beta=1$ corresponds to the available `random'
processes. A path through the parameter space $\cal{S}$ of gamma
models determines a curve
\begin{equation} c:[a,b]\rightarrow {\cal{S}}: t\mapsto
(c_1(t),c_2(t))\end{equation} with tangent vector
$\dot{c}(t)=(\dot{c}_1(t),\dot{c}_2(t))$ and norm $||\dot{c}||$
given via~(\ref{gammametric}) by
\begin{equation} ||\dot{c}(t)||^2=\frac{c_2(t)}{c_1(t)^2} \, \dot{c}_1(t)^2 +
        \left(\psi'(c_2(t))
        -\frac{1}{c_2(t)}\right)\,\dot{c}_2(t)^2.\end{equation}
The information length of the curve is \begin{equation}
L_c(a,b)=\int_a^b||\dot{c}(t)|| \, dt \end{equation} and the curve
corresponding to an underlying Poisson process has $c(t)=(t,1),$
so $t=\mu$ and $\beta=1=constant,$ and the information length is
$\log\frac{b}{a}.$

\vspace{1cm}
{\bf Acknowledgement} The author is grateful to C. Frenk and J. Colberg for comments
during the preparation of this
article.


\begin{thebibliography}{99}
\bibitem{gamran} Khadiga Arwini and C.T.J. Dodson.
Information geometric neighbourhoods of randomness and geometry of
the McKay bivariate gamma 3-manifold. {\em Sankhya: Indian Journal
of Statistics}, 66, 2 (2004) 211-231.

\bibitem{bacc} C. Baccigalupi, L. Amendola and F. Occhionero.
Imprints of primordial voids on the cosmic microwave background
{\em Mon. Not. R. Astr. Soc.} 288, 2
    (1997) 387-96.

\bibitem{benson} A.J. Benson, F. Hoyle, F. Torres and M.J, Vogeley.
LGalaxy voids in cold dark matter universes.
{\em Mon. Not. R. Astr. Soc.} 340 (2003) 160-174.


\bibitem{coles} P. Coles. Understanding recent observations of the large-scale
structure of the universe. {\em Nature} 346 (1990) 446-.

\bibitem{croton04a} D.J. Croton et al. (The 2dFGRS Team).The 2dF Galaxy Redshift Survey:
Higher order galaxy correlation functions. Preprint, arXiv:astro-ph/0401434 v2 23 Aug 2004.

\bibitem{croton04b} D.J. Croton et al. (The 2dFGRS Team). The 2dF Galaxy Redshift Survey:
Voids and hierarchical scaling models. Preprint, arXiv:astro-ph/0401406 v2 23 Aug 2004.

\bibitem{vpf} C.T.J. Dodson. Evolution of the void probability function.
Presented at {\bf Workshop on Statistics of Cosmological Data
Sets}, 8-13 August 1999, Isaac Newton Institute, Cambridge. \\
\verb+http://www.maths.manchester.ac.uk/~kd/PREPRINTS/vpf.ps+

\bibitem{ijtp} C.T.J. Dodson. Spatial statistics
and information geometry for
parametric statistical models of galaxy clustering. {\em Int. J. Theor. Phys.}
38, 10 (1999) 2585-2597.

\bibitem{gsis} C.T.J. Dodson.
Geometry for stochastically inhomogeneous spacetimes. {\em
Nonlinear Analysis}, 47 (2001) 2951-2958.


\bibitem{InfoGeom} C.T.J. Dodson.
Information Geometry. Workshop presentation, CIMAT Mexico 2004,
Santiago de Compostela, Spain 2005. Pages 1-60. \\ {\em
Preprint:}
\verb+http://www.maths.manchester.ac.uk/~kd/PREPRINTS/InfoGeom.pdf+

\bibitem{dodsonsampson} C.T.J. Dodson and W.W. Sampson. Modeling a class of
stochastic porous media. {\em Appl. Math. Lett.} 10, 2 (1997) 87-89.

\bibitem{fairall} A.P. Fairall. {\bf Large-scale structure in the universe}
Wiley-Praxis, Chichester 1998.

\bibitem{hoyle} F. Hoyle and M.S. Vogeley. Voids in the 2dF Galaxy Redshift Survey.
{\em Astrophys. J.} 607 (2004) 751-764.

\bibitem{jaynes} E.T. Jaynes. Information theory and statistical inference.
{\em The Physical Review} 106 (1957) 620-630 and 108 (1957) 171-190.

\bibitem{kauffmannfairall} G. Kauffmann and A.P. Fairall. Voids in the distribution of
galaxies: an assessment of their significance and derivation of a void spectrum.
{\em Mon. Not. R. Astr. Soc.} 248 (1990) 313-324.

\bibitem{lachieze} M. Lachi\'{e}ze-Rey, L.N. Da-Costa and S. Maurogordato.
Void probability function in the Southern
Sky Redshift Survey. {\em Astrophys. J.} 399 (1992) 10-15.

\bibitem{lauritzen} S.L. Lauritzen. Statistical Manifolds. In {\bf
Differential Geometry in Statistical Inference}, Institute of Mathematical
Statistics Lecture Notes, Volume 10, Berkeley 1987, pp 163-218.


\bibitem{piran} T. Piran, M.Lecar, D.S. Goldwirth, L. Nicolaci da Costa
    and G.R. Blumenthal. Limits on the primordial fluctuation
spectrum: void sizes and anisotropy of the cosmic microwave background radiation.
{\em Mon. Not. R. Astr. Soc.} 265, 3 (1993) 681-8.





\bibitem{weygaert} R. van der Weygaert. Quasi-periodicity in deep redshift surveys. {\em Mon. Not. R. Astr. Soc.} 249 (1991) 159-.

\bibitem{weygaerticke} R. van der Weygaert and V. Icke. Fragmenting the
universe II. Voronoi vertices as Abell clusters. {\em Astron. Astrophys.} 213 (1989) 1-9.

\bibitem{white} S.D.M. White. The hierarchy of correlation functions and its relation
to other measures of galaxy clustering. {\em Mon. Not. R. Astr. Soc.} 186,
     (1979) 145-154.

\end{thebibliography}
\end{document}